\documentclass[manuscript]{aastex}

\usepackage{hyperref}
\usepackage{url}
\usepackage{lscape}
\usepackage{longtable}
\usepackage{latexsym}
\usepackage{graphics}
\usepackage{graphicx}
\usepackage{enumitem}
\usepackage{amsmath}
\usepackage{mathptmx}
\usepackage{latexsym}
\usepackage{fix-cm}
\usepackage{bigstrut}
\usepackage{booktabs}

\begin{document}

\title{A new solution of Einstein's field equations in isotropic coordinates}

\author{B. S. Ratanpal\altaffilmark{1}}
\affil{Department of Applied Mathematics, Faculty of Technology \& Engineering, The M. S. University of Baroda, Vadodara - 390 001, India}
\email{bharatratanpal@gmail.com}

\author{Bhavesh Suthar\altaffilmark{2}}
\affil{Department of Applied Mathematics, Faculty of Technology \& Engineering, The M. S. University of Baroda, Vadodara - 390 001, India}
\email{bhaveshsuthar.math@gmail.com}

\begin{abstract}
In this work, an exact solution of Einstein's field equations in isotropic coordinates for anisotropic matter distribution is obtained by considering a particular metric choice of metric potential $g_{rr}$. To check the feasibility of the model, we have investigated all the physical characteristics of a realistic star. It is found that the model is potentially stable, and the adiabatic index is greater than $\frac{4}{3}$. The model have been analysed for Compact star \textbf{4U 1538-52}.
\end{abstract}


\section{Introduction}
The search for exact solutions of Einstein’s field equations on geometrically significant spacetime that satisfies physical constraints has remained the subject of great interest to mathematicians and physicists. Such findings are also crucial in relativistic astrophysics as they enable the distribution of matter in the interior of the stellar objects to be modelled in terms of simple algebraic relations. Due to the non-linear nature of Einstein’s field equations, it becomes difficult to obtain new exact solutions. Moreover, to be a realistic solution, it must satisfy certain conditions for physical acceptability. There are many comprehensive collections of static, spherically symmetric solutions available in the literature.

Since the pioneering work of \cite{Schwarzschild1916}, who obtained the first interior solution describing a uniform density sphere, the modelling of relativistic stars has moved from the regime of toy models to sophisticated, realistic stellar structures. The theoretical investigations of \cite{Ruderman1972} about more realistic stellar models show that the nuclear matter may be locally anisotropic, at very high-density. According to these views, the radial pressure may not be equal to the tangential pressure in massive stellar objects. Some of the exciting research in this regard include the work of \cite{Dev2002}, \cite{Mak2002}, \cite{Mak2002-1}, \cite{Chaisi2006}, \cite{Komathiraj2007}, \cite{Komathiraj2007-1}, \cite{Thirukkanesh2012}, \cite{Maurya2013}. An exact solution of Einstein's field equations for anisotropic fluid distribution on the background of a pseudo-spheroidal spacetime has been obtained by \cite{Tikekar1999}, \cite{Thomas2015} $\&$ \cite{Ratanpal2016}. \cite{Thomas2017} have discussed new exact solutions of Einstein’s field equations by assuming a linear equation of state on the background of a paraboloidal spacetime. \cite{Pandya2019} have reported an exact solution of Einstein's field equation on the background of paraboloidal spacetime using Karmarkar conditions. \cite{Thirukkanesh2008} assumed a linear equation of state to obtain the solution of anisotropic fluid distribution. \cite{Thirukkanesh2012-1} used the polytropic equation of state for developing anisotropic models. \cite{Sharma2013} studied an anisotropic model of superdense stars admitting the quadratic equation of state. Exact solutions of Einstein's field equations for charged anisotropic distribution on the background of pseudo-spheroidal spacetime have been obtained by \cite{Ratanpal2015} $\&$ \cite{Thomas2015-1}. The charged anisotropic star on paraboloidal spacetime has been reported by \cite{Ratanpal2016-1}. A new model of a charged compact star by solving the Einstein-Maxwell field equations was reported by \cite{Ratanpal2017}.

It has been shown by \cite{Nariai1950} that every line element given in a canonical like Schwarzchild coordinate system can easily be transformed into the isotopic coordinate system, though the vice-versa is not always possible. The isotropic line elements treat three spatial dimensions in the same way. \cite{Knutsen1991} has proposed Goldman's method to construct physically valid fluid spheres. \cite{Kuchowicz1971}, \cite{Kuchowicz1971-1}, \cite{Kuchowicz1972}, \cite{Kuchowicz1972-1}, \cite{Kuchowicz1973} showed experimental approaches to solve Einstein’s field equations in isotropic distribution. He has proposed an exact solution for perfect fluid in isotropic coordinates. The generation technique for more new solutions in isotopic distribution was used by \cite{Hajj1986}. \cite{Mak2005} have reduced Einstein’s field equations to two independent Riccati differential equations for which three classes of solutions are obtained by using a matrix transformation in isotropic coordinates.  \cite{Rahman2002} and \cite{Lake2003} also discovered the exact solutions of Einstein’s field equations for spherically symmetric perfect fluid solution in the background of isotopic coordinate. The exact solution of Einstein’s field equations in an isotropic coordinate system have been obtained in numerous work. Some interesting work in this regard include \cite{Ngubelanga2013}, \cite{Murad2014}, \cite{Pant2014}, \cite{Pant2014-1}, \cite{Pant2014-2}, \cite{Newton2015}, \cite{Govender2016}, \cite{Pant2015}, \cite{Pant2015-1}, \cite{Ngubelanga2015}, \cite{Ngubelanga2017}, \cite{Ngubelanga2015-1}, \cite{Bhar2018}.

In this work, the exact solution of Einstein’s field equations in isotropic coordinate system is obtained by choosing a particular ansatz for metric potential $e^ \lambda $. It is observed that the solution is non-singular and satisfies all physical plausibility conditions. We analysed the results for the star 4U 1538-52. The paper has been organised as follows: In section \ref{sec:2}, Einstein field equations for anisotropic distribution of matter in isotropic coordinates have been reported by considering particular ansatz for metric potential $e^ \lambda $. Also, the solution of Einstein's field equations have been discussed in this section. Section \ref{sec:3}, contains a physical analysis of the model. Finally, in section \ref{sec:4}, we conclude the work.

\section{Solution of Einstein's Field Equations}
\label{sec:2}
We consider a static and spherically symmetric spacetime metric in isotropic coordinates
\begin{equation} \label{IMetric}
    ds^{2}=-A^2(r)dt^2+B^2(r)[dr^2+r^{2}\left(d\theta^{2}+\sin^{2}\theta d\phi^{2} \right)],
\end{equation}
where A(r) and B(r) are arbitrary functions of radial coordinate r.\\
The matter distribution of the interior of this stellar object is described by the energy-momentum tensor
\begin{equation} \label{EMTensor}
    T_{ij}= diag(-\rho,p_{r},p_{t},p_{t}),
\end{equation}
where $\rho$, $p_{r}$ and $p_{t}$ denote the energy density, radial pressure and tangential pressure respectively, and $\Delta = p_{t}-p_{r}  $ represents anisotropy.
The radial pressure and the tangential pressure are measured relative to the comoving fluid four-velocity
\begin{equation} \label{UFV}
    u^{a}=\frac 1{A} \delta^{a}.
\end{equation}
The Einstein's field equations corresponding to the spacetime metric (\ref{IMetric}) and the energy-momentum tensor (\ref{EMTensor}) take the form
\begin{equation} \label{Rho1}
    8\pi\rho=\frac{-1}{B^{2}}\Bigg[2\frac{B^{''}}{B}-\frac{{B^{'}}^{2}}{{B}^{2}}+\frac{4}{r}\frac{B^{'}}{B}\Bigg],
\end{equation}
\begin{equation} \label{Pr1}
    8\pi p_{r}=\frac{1}{B^2}\Bigg[\frac{{B^{'}}^{2}}{B^{2}}+2\frac{A^{'}}{A}\frac{B^{'}}{B}+\frac{2}{r}\Bigg(\frac{A^{'}}{A}+\frac{B^{'}}{B}\Bigg)\Bigg],
\end{equation}
and
\begin{equation} \label{Pt1}
    8\pi p_{t}=\frac{1}{B^2}\Bigg[\frac{A^{''}}{A}+\frac{B^{''}}{B}-\frac{{B^{'}}^{2}}{B^2}+\frac{1}{r}\Bigg(\frac{A^{'}}{A}+\frac{B^{'}}{B}\Bigg)\Bigg],
\end{equation}
where prime (') denote differentiation with respect to the radial coordinate r.\\
To obtain solution of above system of equations (\ref{Rho1})-(\ref{Pt1}), we use relativistic units in which $8\pi G=1$ and speed of light $c=1$. From (\ref{Pr1}) and (\ref{Pt1}), we get the following differential equation
\begin{equation} \label{Anisotorpy1}
    \Delta=p_{t}-p_{r}=\frac{1}{B^{2}}\Bigg[\frac{A^{''}}{A}+\frac{B^{''}}{B}-2\frac{{B^{''}}^{2}}{B^{2}}-2\frac{A^{'}}{A}\frac{B^{'}}{B}-\frac{1}{r}\Bigg(\frac{A^{'}}{A}+\frac{B^{'}}{B}\Bigg)\Bigg].
\end{equation}

To solve the above system of equations, various choices can be made for the gravitational potential B(r). Regardless, the choice of B(r) must be physically reasonable for a realistic stellar model.
Following \cite{Govender2015}, we choose B(r) as
\begin{equation} \label{B}
    B(r)=\frac{a}{\sqrt{1+br^{2}}},
\end{equation}
where a and b are constants. The above chosen gravitation potential B(r) is regular at the center. \cite{Einstein1917}, and \cite{Sitter1917} utilised the same ansatz for B(r) in curvature coordinates. Recently \cite{Govender2015}, \cite{Thirukkanesh2015}, and \cite{Bhar2018} used the same expression of B(r) in isotropic coordinates. By utilising this choice of B(r), we obtained the energy density ($\rho$), radial pressure ($p_{r}$) and tangential pressure ($p_{t}$) as
\begin{equation} \label{Rho2}
    8\pi\rho=\frac{b(6+br^{2})}{a^{2}(1+br^{2})},
\end{equation}
\begin{equation} \label{pr2}
    8\pi p_{r}=\frac{2A^{'}\left(1+br^{2}\right)-Abr\left(2+br^{2}\right)}{Aa^{2}r\left(1+br^{2}\right)},
\end{equation}
and
\begin{equation} \label{pt2}
    8\pi p_{t}=\frac{{A}^{''}r{(1+br^{2})}^{2}+{A}^{'}{(1+br^{2})}^{2}-2Abr}{A{a}^{2}r(1+br^{2})}.
\end{equation}
Therefore, the measure of anisotropy ($\Delta$) takes the form
\begin{equation} \label{Deltaeqn}
    \Delta=\frac{{A}^{''}r{\left(1+br^{2}\right)}^{2}+{A}^{'}\left(1+br^{2}\right)\left(-1+br^{2}\right)+Ab^{2}r^{3}}{Aa^{2}r\left(1+br^{2}\right)}.
\end{equation}
The equation (\ref{Deltaeqn}) gives
\begin{equation} \label{Eq10}
    A^{''}+\frac{A^{'}(br^{2}-1)}{r(1+br^{2})}+\frac{(b^{2}r^{2}-\Delta a^{2}(1+br^{2}))}{(1+br^{2})^{2}}A=0.
\end{equation}
To obtain a non-singular solution, we make a particular choice of anisotropy
\begin{equation}\label{Anisotropy2}
    \Delta=\frac{b^{2}r^{2}}{a^{2}(1+br^{2})}.
\end{equation}
It can be seen that $\Delta=0$ at $r=0$. Combining equation  (\ref{Anisotropy2}) in (\ref{Eq10}), we get
\begin{equation}\label{Eq15}
    {A}^{''}+\frac{{A}^{'}\left(br^{2}-1\right)}{r\left(1+br^{2}\right)}=0.
\end{equation}
The solution of equation (\ref{Eq15}) is given by
\begin{equation} \label{A1}
    A=\frac{C\log(1+br^{2})+2bD}{2b},
\end{equation}
where C and D are constants of integration.\\
Substituting equation (\ref{A1}) in equation (\ref{pr2}) and equation (\ref{pt2})
\begin{equation}
    8\pi p_{r}=-\frac{b\left[-4C+2bD\left(2+br^{2}\right)+C\left(2+br^{2}\right)\log\left(1+br^{2}\right)\right]}{a^{2}\left(1+br^{2}\right)\left[2bD+C\log\left(1+br^{2}\right)\right]},
\end{equation}
\begin{equation}
    8\pi p_{t}=-\frac{2b\left[-2C+2bD+C\log(1+br^{2})\right]}{a^{2}(1+br^{2})\left[2bD+C\log(1+br^{2})\right]}.
\end{equation}
Hence, the spacetime metric (\ref{IMetric}) takes the form
\begin{equation} \label{Metric2}
    ds^{2}=-\frac{\big[2bD+C\log(1+b {r}^{2})\big]^{2}}{4b^{2}}dt^{2}+\frac{a^{2}}{(1+br^{2})}\big[dr^{2}+r^{2}{d\theta}^{2}+r^{2}\sin^{2}{\theta}{d\phi}^{2}\big].
\end{equation}
For a physically acceptable model, the interior spacetime metric (\ref{Metric2}) should continuous match with the Schwarzchild exterior spacetime metric in isotropic coordinates
\begin{equation}\label{EMetric}
	ds^{2}=-\frac{\Big(1-\frac{M}{2r} \Big)^{2}}{\Big(1+\frac{M}{2r} \Big)^{2}}dt^{2}+\left(1+\frac{M}{2r} \right)^{4}dr^{2}+r^{2}\left(d\theta^{2}+\sin^{2}\theta d\phi^{2} \right),
\end{equation}
across the boundary $r=R$ of the star together with the condition that the radial pressure should vanish at the surface $\left(p_{r}\left(r=R\right)=0\right)$.
Comparing the relevant coefficient across the boundary R then yields
\begin{equation} \label{M1}
    M=2R\left[\frac{\sqrt{a}}{\left(1+bR^{2}\right)^\frac{1}{4}}-1\right],
\end{equation}
\begin{equation} \label{C1}
    C=\frac{b\left(2+bR^{2}\right)\left[-\sqrt{a}+2\left(1+bR^{2}\right)^\frac{1}{4}\right]}{2\sqrt{a}},
\end{equation}
and
\begin{equation} \label{D1}
    D=\frac{\left[2+\left(1+bR^{2}\right)^\frac{1}{4}-\sqrt{a}\right]\left[4-\left(2+bR^{2}\right)\log \left(1+bR^{2}\right)\right]}{4\sqrt{a}}.
\end{equation}
Using equations (\ref{C1}) and (\ref{D1}) in equation (\ref{A1}), we get the value of A as follows
\begin{equation} \label{A2}
    A=\frac{\left[-\sqrt{a}+2\left(1+bR^{2}\right)^\frac{1}{4}\right]\left[4+\left(2+bR^{2}\right)\log\left(1+br^{2}\right)-\left(2+bR^{2}\right)\log\left(1+bR^{2}\right)\right]}{4\sqrt{a}}.
\end{equation}
Subsequently, the expressions for energy density, radial pressure and tangential pressure can be given as
\begin{equation} \label{Rho3}
    \rho=\frac{b\left(6+br^{2}\right)}{a^{2}\left(1+br^{2}\right)},
\end{equation}
\footnotesize
\begin{equation} \label{Pr}
    p_{r}=\frac{b\left[4b\left(-r^{2}+R^{2}\right)-\left(4+b^{2}r^{2}R^{2}+2b\left(r^{2}+R^{2}\right)\right)\log\left(1+br^{2}\right)+\left(4+b^{2}r^{2}R^{2}+2b\left(r^{2}+R^{2}\right)\right)\log\left(1+bR^{2}\right)\right]}{a^{2}\left(1+br^{2}\right)\left[4+\left(2+bR^{2}\right)\log\left(1+br^{2}\right)-\left(2+bR^{2}\right)\log\left(1+bR^{2}\right)\right]},
\end{equation}
\normalsize
and
\begin{equation} \label{Pt}
    p_{t}=\frac{2b\left[2bR^{2}-\left(2+bR^{2}\right)\log\left(1+br^{2}\right)+\left(2+bR^{2}\right)\log\left(1+bR^{2}\right)\right]}{a^{2}\left(1+br^{2}\right)\left[4+\left(2+bR^{2}\right)\log\left(1+br^{2}\right)-\left(2+bR^{2}\right)\log\left(1+bR^{2}\right)\right]}.
\end{equation}
We analyze these physical entities in the next section.

\section{Physical Analysis}
\label{sec:3}
\begin{figure}
    \includegraphics[scale = 1.25]{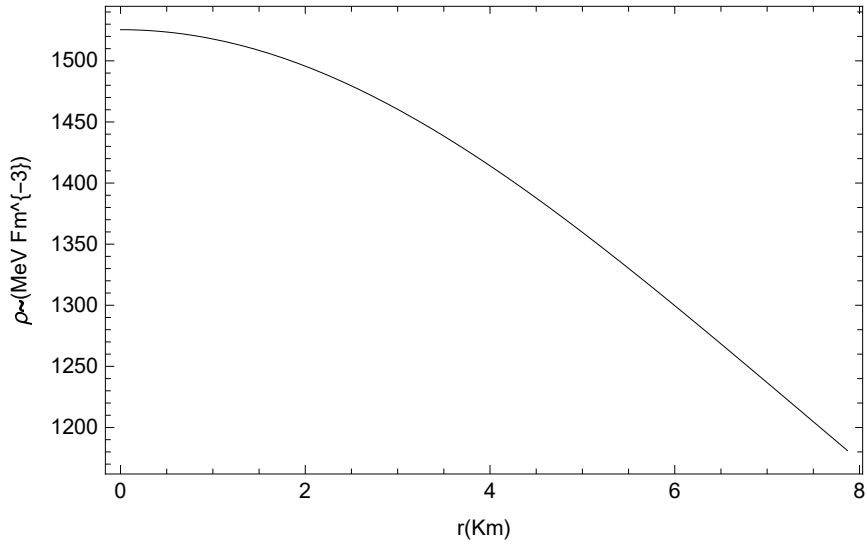}
    \caption{Variation of energy density ($\rho$) with respect to radius (r) inside a stellar interior for a star 4U 1538-52 with R = 7.87 km for a = 0.843, and b = 0.006.}
    \label{fig:Figure. 1}
\end{figure}

\begin{figure}
    \includegraphics[scale = 1.25]{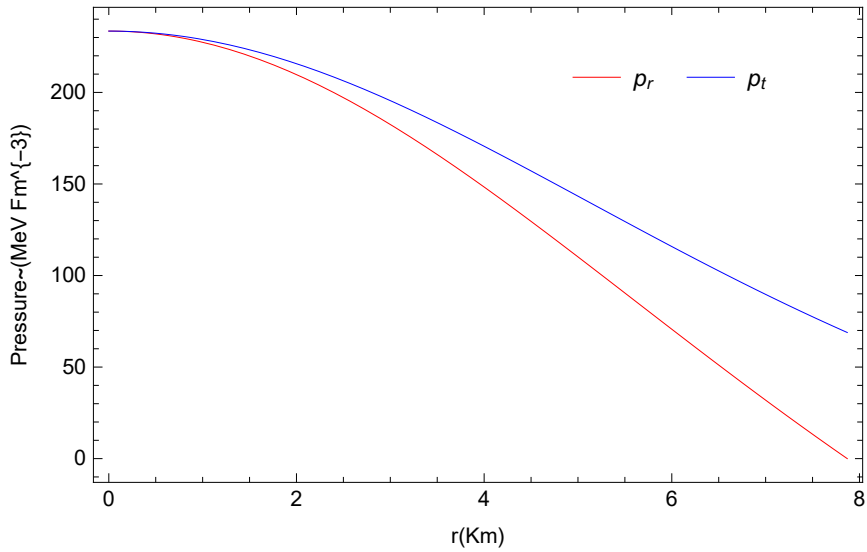}
    \caption{Variation of pressures ($p_{r}, and p_{t}$) with respect to radius (r) inside a stellar interior for a star 4U 1538-52 with R = 7.87 km for a = 0.843, and b = 0.006.}
    \label{fig:Figure. 2}
\end{figure}

\begin{figure}
    \includegraphics[scale = 1.25]{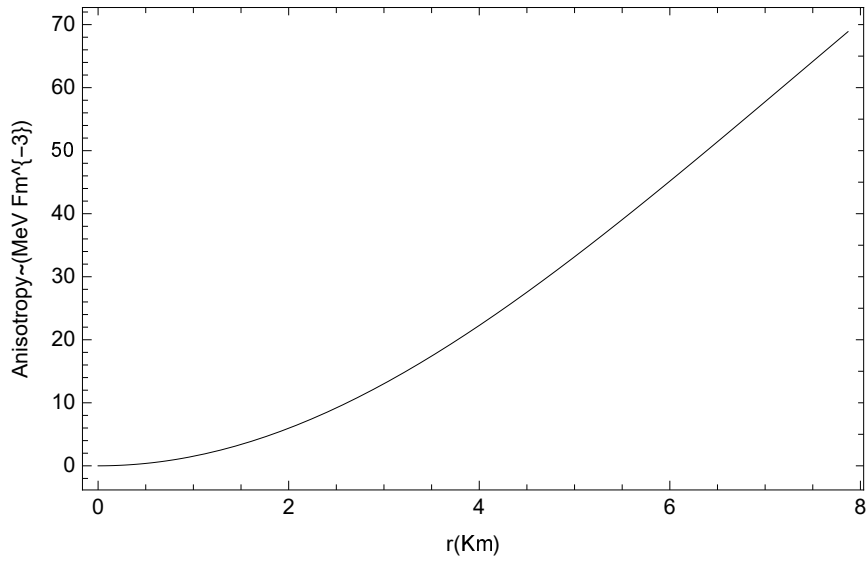}
    \caption{Measure of Anisotropy ($\Delta$) against radial coordinate r for a star 4U 1538-52 with R = 7.87 km for a = 0.843, and b = 0.006.}
    \label{fig:Figure. 3}
\end{figure}

\begin{figure}
    \includegraphics[scale = 1.25]{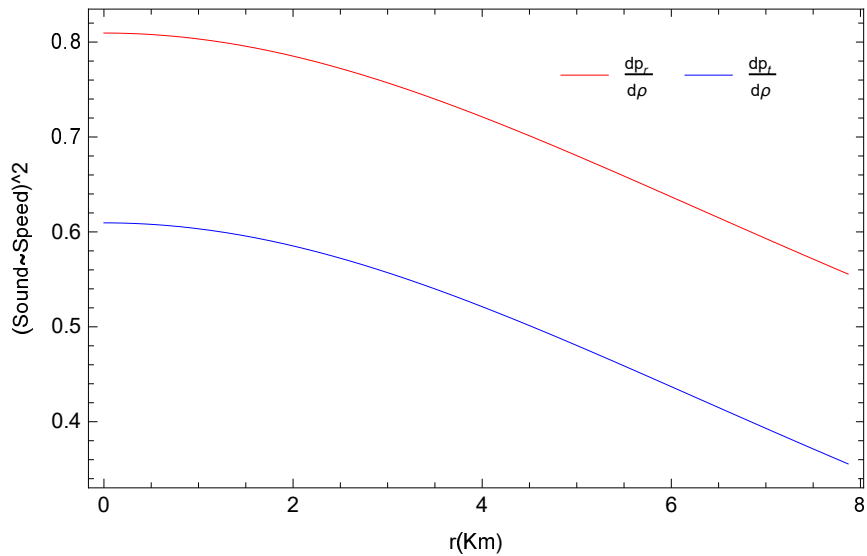}
    \caption{Squared Sound Speed against radial coordinate r for a star 4U 1538-52 with R = 7.87 km for a = 0.843, and b = 0.006.}
    \label{fig:Figure. 4}
\end{figure}

\begin{figure}
    \includegraphics[scale = 1.25]{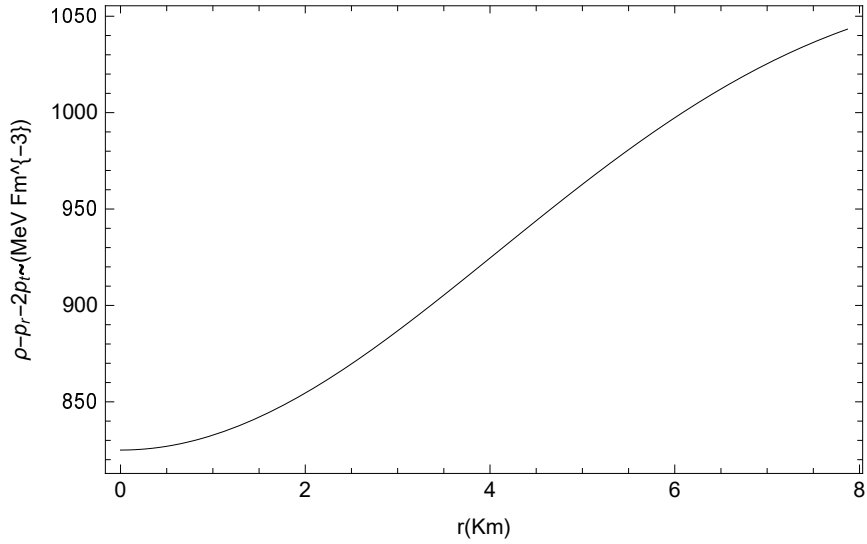}
    \caption{Strong Energy Condition ($\rho-p_{r}-2p_{t}$) against radial coordinate r for a star 4U 1538-52 with R = 7.87 km for a = 0.843, and b = 0.006.}
    \label{fig:Figure. 5}
\end{figure}

\begin{figure}
    \includegraphics[scale = 1.25]{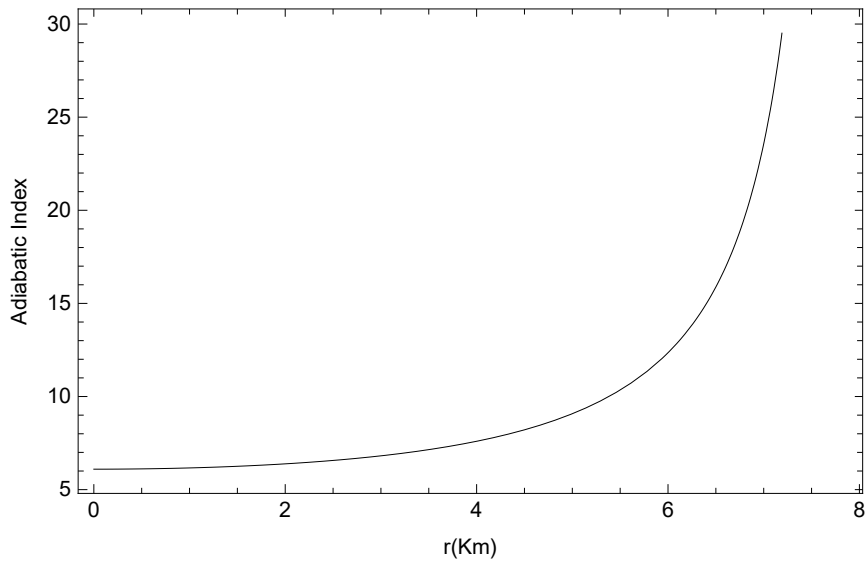}
    \caption{Adiabatic Index ($\Gamma$) against radial coordinate r for a star 4U 1538-52 with R = 7.87 km for a = 0.843, and b = 0.006.}
    \label{fig:Figure. 6}
\end{figure}

The model should satisfy the following physical plausibility conditions proposed by \cite{Knusten1988}.
\begin{itemize}
	\item[(i)]  Energy density ($\rho$), radial pressure ($p_{r}$) and tangential pressure ($p_{t}$) all should be positive throughout the distribution.
	\item[(ii)] They should be decreasing throughout the distribution.\\
	i.e., $\frac{d\rho}{dr}<0$, $\frac{dp_{r}}{dr}<0$, $\frac{dp_{t}}{dr}<0$ for $0\leq r\leq R$.
	\item[(iii)] Causality condition should be satisfied inside the stellar interior.\\
	i.e., $0<\frac{dp_{r}}{d\rho}<1$, $0<\frac{dp_{t}}{d\rho}<1$ for $0\leq r\leq R$.
	\item[(iv)] The solution should satisfy the strong energy condition defined by $\rho-p_{r}-2p_{t}\geq 0$ for $0\leq r \leq R$.
\end{itemize}
In order to have a detailed analysis of various physical conditions, we have considered the radius of a particular star proposed by \cite{Gangopadhyay}. We have examined the results for particular star 4U 1538-52 having radius $R=7.87 km$ by taking $a=0.843$ $km^{-2}$ and $b=0.006$ $km^{-2}$. Figure \ref{fig:Figure. 1} and Figure \ref{fig:Figure. 2} demonstrate the variation of energy density and pressure from the centre to the boundary of the star. It can be observed that energy density and pressure are monotonically decreasing functions of r. The variation of anisotropy ($\Delta$) is plotted in Figure \ref{fig:Figure. 3}. The graph clearly indicates that the anisotropy ($\Delta$) is zero at the centre and $\Delta>0$, which is a desirable characteristic of the compact object given by \cite{Gokhroo}. The variation of a square of sound speed is displayed in Figure \ref{fig:Figure. 4}. It is noticed that  the square of sound speed lies between [0,1]. Figure \ref{fig:Figure. 5} reflects that the strong energy condition is satisfied throughout the distribution. The stability of stellar configuration is depends on adiabatic index ($\Gamma$). Figure \ref{fig:Figure. 6} clearly indicates that the adiabatic index ($\Gamma$) is greater than $\frac{4}{3}$ throughout the configuration. The present model is in good agreement with the star 4U 1538-52 and satisfies all physical acceptability conditions.

\section{Conclusion}
\label{sec:4}
To construct a physically acceptable stellar model, many authors imposed an equation of state which relates the pressure to the density of the star, i.e., $p=p(\rho)$. \cite{Ngubelanga2015-2} used linear equation of state in isotropic coordinates for physically viable relativistic models of compact stars. \cite{Ngubelanga2015} have used the quadratic equation of state for obtaining the solution of anisotropic distribution. \cite{Ngubelanga2015-1} have used the polytropic equation of state for generating solutions for a relativistic star. \cite{Bhar2018} employed Chaplygin equation of state to obtain the solution of Einstein's field equations in isotropic coordinates. \cite{Malaver2013} used Van der Waals modified equation of state to construct a physically reasonable model. Since at ultra high density, the equation of state may not be known hence we constructed the model in which the equation of state is not precisely known. We have obtained the singularity-free solution of Einstein's field equations in isotropic coordinates. We have examined the physical viability of our model by investing particular model of pulsar 4U 1538-52. From the figures, it is clear that the model satisfies all the physical plausibility conditions. Moreover, it is found that $\Gamma>\frac{4}{3}$ also model is potentially stable under the criteria given by \cite{Ratnapal2020}.



\begin{thebibliography}{00}
\bibitem[\protect \citeauthoryear{Schwarzchild}{1916}]{Schwarzschild1916} Schwarzchild K., {\it Sitz. P. Akad. Wiss. Berlin, Kl. Math. Phys.} {\bf1} (1916) 189.
\bibitem[\protect \citeauthoryear{Ruderman}{1972}]{Ruderman1972} Ruderman M., {\it Ann. Rev. Astron. Astrophys.} {\bf10} (1972) 427.
\bibitem[\protect \citeauthoryear{Dev and Gleiser}{2002}]{Dev2002} Dev K. and Gleiser M., {\it Gen. Rel. Grav.} {\bf34} (2002) 1793.
\bibitem[\protect \citeauthoryear{Mak {\em et al}}{2002}]{Mak2002} Mak M. K., Dobson P. N. and Harko T., {\it Int. J. Mod. Phys. D} {\bf11} (2002) 207.
\bibitem[\protect \citeauthoryear{Mak and Harko}{2002}]{Mak2002-1} Mak M. K. and Harko T., {\it hin. J. Astron. Astrophys.} {\bf2} (2002) 248.
\bibitem[\protect \citeauthoryear{Chaisi and Maharaj}{2006}]{Chaisi2006} Chaisi M. and Maharaj S. D., {\it J. Phys.} {\bf66} (2006) 609.
\bibitem[\protect \citeauthoryear{Komathiraj and Maharaj}{2007}]{Komathiraj2007} Komathiraj K. and Maharaj S. D., {\it J. Math. Phys.} {\bf48} (2007) 042501.
\bibitem[\protect \citeauthoryear{Komathiraj and Maharaj}{2007}]{Komathiraj2007-1} Komathiraj K. and Maharaj S. D., {\it Int. J. Mod. Phys. D} {\bf16} (2007) 1803.
\bibitem[\protect \citeauthoryear{Thirukkanesh and Ragel}{2012}]{Thirukkanesh2012} Thirukkanesh S. and Ragel F. C., {\it Pramana J. Phys.} {\bf78} (2012) 687.
\bibitem[\protect \citeauthoryear{Maurya and Gupta}{2013}]{Maurya2013} Maurya S. K. and Gupta Y. K., {\it Astrophys. Space Sci.} {\bf344} (2013) 243.
\bibitem[\protect \citeauthoryear{Tikekar and Thomas}{1999}]{Tikekar1999} Tikekar R. and Thomas V. O., {\it Pramana J. Phys.} {\bf52} (1999) 237.
\bibitem[\protect \citeauthoryear{Thomas and Pandya}{2015}]{Thomas2015} V. O. Thomas and D. M. Pandya, {\it Astrophys. Space Sci.} {\bf360} (2015) 59.
\bibitem[\protect \citeauthoryear{Ratanpal {\em et al}}{2016}]{Ratanpal2016} Ratanpal  B. S., Thomas V. O. and Pandya D. M., {\it Astrophys. Space Sci.} {\bf361} (2016) 65.
\bibitem[\protect \citeauthoryear{Thomas and Pandya}{2017}]{Thomas2017} Thomas V. O. and Pandya D. M., {\it Eur. Phys. J. A} {\bf53} (2017) 120.
\bibitem[\protect \citeauthoryear{Pandya and Thomas}{2019}]{Pandya2019} Pandya D. M. and Thomas V. O., {\it Can. J. Phys.} {\bf97} (2019) 337.
\bibitem[\protect \citeauthoryear{Thirukkanesh and Maharaj}{2008}]{Thirukkanesh2008} Thirukkanesh S. and Maharaj S. D., {\it Class. Quantum Grav.} {\bf25} (2008) 235001.
\bibitem[\protect \citeauthoryear{Thirukkanesh and Ragel}{2012}]{Thirukkanesh2012-1} Thirukkanesh S. and Ragel  F. C., {\it Pramana J. Phys.} {\bf78} (2012) 687.
\bibitem[\protect \citeauthoryear{Sharma and Ratanpal}{2013}]{Sharma2013} Sharma R. and Ratanpal B. S., {\it Int. J. Mod. Phys. D} {\bf22} (2013) 1350074.
\bibitem[\protect \citeauthoryear{Ratanpal {\em et al}}{2015}]{Ratanpal2015} Ratanpal  B. S., Thomas V. O. and Pandya D. M., {\it Astrophys. Space Sci.} {\bf360} (2015) 53.
\bibitem[\protect \citeauthoryear{Thomas and Pandya}{2015}]{Thomas2015-1} Thomas V. O. and Pandya D. M., {\it Astrophys. Space Sci.} {\bf360} (2015) 39.
\bibitem[\protect \citeauthoryear{Ratanpal and Sharma}{2016}]{Ratanpal2016-1} Ratanpal B. S. and Sharma J., {\it Pramana J. Phys.} {\bf86} (2016) 527.
\bibitem[\protect \citeauthoryear{Ratanpal and Bhar}{2017}]{Ratanpal2017} Ratanpal B. S. and Bhar P., {\it Phys. Astron. Int. J.} {\bf1} (2017) 151.
\bibitem[\protect \citeauthoryear{Nariai}{1950}]{Nariai1950} Nariai H., {\it Sci. R. Tóhoku Uni.} {\bf34} (1950) 160.
\bibitem[\protect \citeauthoryear{Knutsen}{1991}]{Knutsen1991} Knutsen H., {\it Gen. Rel. Grav.} {\bf23} (1991) 843.
\bibitem[\protect \citeauthoryear{Kuchowicz}{1971}]{Kuchowicz1971} Kuchowicz B., {\it Phys. Let. A} {\bf35} (1971) 223.
\bibitem[\protect \citeauthoryear{Kuchowicz}{1971}]{Kuchowicz1971-1} Kuchowicz B., {\it Acta Physica Polonica} {\bf B2} (1971) 657.
\bibitem[\protect \citeauthoryear{Kuchowicz}{1972}]{Kuchowicz1972} Kuchowicz B., {\it Phys. Let. A} {\bf38} (1972) 369.
\bibitem[\protect \citeauthoryear{Kuchowicz}{1972}]{Kuchowicz1972-1} Kuchowicz B., {\it Acta Physica Polonica} {\bf B3} (1972) 209.
\bibitem[\protect \citeauthoryear{Kuchowicz}{1973}]{Kuchowicz1973} Kuchowicz B., {\it Acta Physica Polonica} {\bf B4} (1973) 415.
\bibitem[\protect \citeauthoryear{Hajj-Boutros}{1986}]{Hajj1986} Hajj-Boutros J., {\it J. Math. Phys.} {\bf27} (1986) 1363.
\bibitem[\protect \citeauthoryear{Mak and Harko}{2005}]{Mak2005} Mak M. K. and Harko T., {\it PRAMANA J. Phys.} {\bf65} (2005) 185.
\bibitem[\protect \citeauthoryear{Rahman and Visser}{2002}]{Rahman2002} Rahman S. and Visser M., {\it Class. Quantum Grav.} {\bf19} (2002) 935.
\bibitem[\protect \citeauthoryear{Lake K.}{2003}]{Lake2003} Lake K., {\it Phys. Rev. D} {\bf67} (2003) 104015.
\bibitem[\protect \citeauthoryear{Ngubelanga and Maharaj}{2013}]{Ngubelanga2013} Ngubelanga S. A. and Maharaj S. D., {\it Adv. Math. Phys.} {\bf2013} (2013) 1.
\bibitem[\protect \citeauthoryear{Murad and Pant}{2014}]{Murad2014} Murad M. H. and Pant N., {\it Astrophys. Space Sci.} {\bf350} (2014) 349.
\bibitem[\protect \citeauthoryear{Pant}{2014}]{Pant2014} Pant N., {\it Int. J. Eng. Innov. Tech.} {\bf4} (2014) 78.
\bibitem[\protect \citeauthoryear{Pant {\em et al}}{2014}]{Pant2014-1} Pant N., Pradhan N. and Murad M. H., {\it Astrophys. Space Sci.} {\bf352} (2014) 135.
\bibitem[\protect \citeauthoryear{Pant {\em et al}}{2015}]{Pant2014-2} Pant N., Pradhan N. and Malaver M., {\it Int. J. Astrophys. Space Sci.} {\bf3} (2015) 1.
\bibitem[\protect \citeauthoryear{Singh {\em et al}}{2015}]{Newton2015} Singh K., Pradhan N. and Malaver M., {\it Int. J. Astrophys. Space Sci.} {\bf3} (2015) 13.
\bibitem[\protect \citeauthoryear{Govender {\em et al}}{2016}]{Govender2016} Govender M., Bogadi R., Sharma R. and Das S., {\it Astrophys. Space Sci.} {\bf361} (2016) 1.
\bibitem[\protect \citeauthoryear{Pant {\em et al}}{2015}]{Pant2015} Pant N., Pradhan N. and Murad M. H., {\it Astrophys. Space Sci.} {\bf355} (2015) 137.
\bibitem[\protect \citeauthoryear{Pant {\em et al}}{2015}]{Pant2015-1} Pant N., Pradhan N., Murad M. H. and Malaver M., {\it A. J. Sci. Tech.} {\bf2} (2015) 43.
\bibitem[\protect \citeauthoryear{Ngubelanga {\em et al}}{2015}]{Ngubelanga2015} Ngubelanga S. A., Maharaj S. D. and Ray S., {\it Astrophys. Space Sci.} {\bf357} (2015) 74.
\bibitem[\protect \citeauthoryear{Ngubelanga and Maharaj}{2017}]{Ngubelanga2017} Ngubelanga S. A. and Maharaj S. D., {\it Astrophys. Space Sci.} {\bf362} (2017) 43.
\bibitem[\protect \citeauthoryear{Ngubelanga and Maharaj}{2015}]{Ngubelanga2015-1} Ngubelanga S. A. and Maharaj S. D., {\it Eur. Phys. J. Plus} {\bf130} (2015) 211.
\bibitem[\protect \citeauthoryear{Bhar {\em et al}}{2018}]{Bhar2018} Bhar P., Govender M. and Sharma R., {\it Pramana J. Phys.} {\bf90} (2018) 5.
\bibitem[\protect \citeauthoryear{Govender and Thirukkanesh}{2015}]{Govender2015} Govender M. and Thirukkanesh S., {\it Astrophys. Space Sci.} {\bf358} (2015) 39.
\bibitem[\protect \citeauthoryear{A. Einstein}{1917}]{Einstein1917} Einstein A., {\it Sitz. Preuss. Akad. Wiss, Berlin Math.Phys.} {\bf1917} (1917) 142.
\bibitem[\protect \citeauthoryear{de Sitter}{1917}]{Sitter1917} de Sitter W., {\it Proc. R. Acad. Amst.} {\bf19} (1917) 1217.
\bibitem[\protect \citeauthoryear{Thirukkanesh {\em et al}}{2015}]{Thirukkanesh2015} Thirukkanesh S., Govender M. and Lortan D. B., {\it Int. J. Mod. Phys. D} {\bf24} (2015) 1550002.
\bibitem[\protect \citeauthoryear{Ngubelanga {\em et al}}{2015}]{Ngubelanga2015-2} Ngubelanga S. A., Maharaj S. D. and Ray S., {\it Astrophys. Space Sci.} {\bf357} (2015) 40.
\bibitem[\protect \citeauthoryear{Knusten}{1988}]{Knusten1988} Knusten H., {\it Astron. Nachr.} {\bf309} (1988) 263.
\bibitem[\protect \citeauthoryear{Gangopadhyay}{2013}]{Gangopadhyay} Gangopadhyay T., Ray S., Li X.D., Dey J. and Dey M., {\it Mon.Not.Roy.Astron.Soc.} {\bf431} (2013) 3216.
\bibitem[\protect \citeauthoryear{Gokhroo and Mehra}{1994}]{Gokhroo} Gokhroo M. K. and Mehra A. L., {\it General Relativity and Gravitation} {\bf26} (1994) 75.
\bibitem[\protect \citeauthoryear{Malaver}{2013}]{Malaver2013} Malaver M., {\it World Applied Programming} {\bf3} (2013) 309.
\bibitem[\protect \citeauthoryear{Ratanpal}{2020}]{Ratnapal2020} Ratanpal B. S., {\it IOP Sci. Notes} {\bf1} (2020) 025207.
\end{thebibliography}
\end{document}